\documentclass{article}
\usepackage{amssymb}
\usepackage{amsmath}
\begin{document}
\title{\textbf{QSES´s and the quantum jump}}
\author{D. Salgado\footnote{E-mail: david.salgado@uam.es}\hspace{0.5em} \& J.L. S\'{a}nchez-G\'{o}mez\footnote{E-mail:jl.sanchezgomez@uam.es}\\
Dpto. F\'{\i}sica Te\'{o}rica, Universidad Aut\'{o}noma de Madrid, Spain}
\maketitle
\begin{abstract}
The stochastic methods in Hilbert space have been used both from a fundamental \cite{GPR} and a practical \cite{Plen} point of view. The result we report here concerns only the idea of applying these methods to model the evolution of quantum systems and does not enter into the question of their fundamental or practical status. It can be easily stated as follows: Once a quantum stochastic evolution scheme is assumed, the incompatibility between the Markov property and the notion of quantum jump is rapidly established.
\end{abstract}

\vspace{1cm}

The paper \cite{DS} is devoted to mathematically prove this result, for which we put forward the previous necessary definitions and theorems.

\bigskip

The mathematical framework is the well-known Hilbert-space-valued stochastic processes formalism, which hereafter we will call Quantum Stochastic Evolution Schemes (QSES's). The physical idea behind this assumption is rather clear: a quantum system (thus described by a vector belonging to a Hilbert space) evolves in a nondeterministic fashion, i.e. only the probability of evolution towards different states can be calculated.

\bigskip

The key tool is the transition probability in a time $t$ from an initial state $\phi$ to a final state $\psi$, in which time homogeneity is implicitly assumed:              
\begin{equation}
P(t;\phi,\psi)\equiv P(0,\phi,t,\psi)=P(u,\phi,t+u,\psi)
\end{equation}
With this tool the Markov property is straightforwardly stated as\footnote{From a strict mathematical point of view this is not the Markov property, but an immediate consequence of it, namely the Chapman-Kolmogorov equation. We do not care about these refinements.} 
\begin{equation}
\int_{\mathcal{H}}P(t;\phi,\psi)P(s;\psi,\varphi)\mu(d\psi)=P(t+s;\phi,\varphi)\quad \forall\phi,\varphi\in\mathcal{H},\forall t,s\in\mathbb{R}^{+}
\end{equation}
The necessary result is the following translation to the QSES language of a well-known theorem in time-continuous Markov chain theory \cite{Chung} 
\begin{list}{
            }
\item {\itshape Let $P(\cdot;\varphi,\psi)$ be a stochastic transition matrix corresponding to a markovian QSES. Then $P(\cdot;\varphi,\psi)$ is continuous in $(0,\infty)$ for all $\varphi,\psi\in\mathcal{H}$ if and only if the following limit exists
\begin{equation}
\lim_{t\to0^{+}}P(t;\phi,\psi)=g(\phi,\psi)
\end{equation}}  
\end{list}
The utility of this result rests on the possibility of checking the continuity of a stochastic matrix at every point through the study of a simple limit at the origin. The only assumed hypothesis is the Markov property.

\bigskip

The main physical hypothesis we are analysing is the quantum jump, one of the most controversial aspects of quantum theory. Using the mathematical language assumed, we state that a QSES reflects the quantum jump if its transition probability satisfies:
\begin{equation}
P(t;\phi,\psi)=\left\{\begin{array}{l}
                     \left\{\begin{array}{ll}
                           1 & \textrm{  if  } 0\leq t\leq t' \textrm{  and  } \psi=U(t)\phi\\
                           0 & \textrm{  if  } 0\leq t\leq t' \textrm{
and } \psi\neq U(t)\phi
                           \end{array}\right.\\
                       \\
                      h(\phi,\psi)(\neq 0) \textrm{  if  } t>t' \textrm{  and it is possible that  }\\
                      \qquad ||\psi-\phi||>\epsilon \textrm{  for some  } \epsilon>0
                      \end{array}\right.
\end{equation}
where $U(t)$ is the usual quantum-mechanical evolution operator.

\bigskip

Notice that $h(\varphi,\psi)=|(\varphi,\psi)|^{2}$ should be expected in order to reproduce the reduction postulate. An excellent framework in which understand the quantum jump is P. Mittelstaedt's analysis of the measurement process \cite{Mitt}, which clearly displays the discontinuity of the state collapse at the instant of objectification and reading of the value of an observable, a feature we are trying to confront with the Markovianity within the QSES's framework.  

\bigskip

The previous theorem provides a very adequate frame to confront the Markovian QSES and the quantum jump hypothesis. We in no case adopt a priori attitudes about the nature of the quantum collapse, just try to enlighten its possible compatibility with the mathematical nature assumed to represent quantum systems.

\bigskip

Once we keep in mind the previous mathematical result, it is easy to prove the following theorem

\begin{list}{\settowidth{\labelwidth}{marg}
             \setlength{\leftmargin}{\labelwidth}
             \setlength{\rightmargin}{\labelwidth}}
\item {\itshape Let $S$ be a quantum system described by a time-homogeneous QSES. If $S$ is subjected to quantum jumps, then its QSES is non-Markovian.}\\
\end{list}

It is straightforward to convince oneself that the central idea of the proof is the existence of the limit at the origin of time in order to apply the mentioned theorem. Though apparently trivial, the question of the existence of such a limit, which in the general theory of Markov chains is called the {\itshape standard condition}, deserves some careful attention which is to be paid in the following section.

\bigskip

Physically the notion of standarization is rather clear: the state of a quantum system does not change if time has hardly elapsed.
In orthodox quantum mechanics the standard condition is automatically fulfilled by the imposition of the initial condition $U(t_{0},t_{0})=I$. Furthermore, standarization appears as natural assumptions in more general frameworks (cf. \cite{Misra},\cite{Dav}) as a hypothesis assumed ``on physical grounds'' in the study of the quantum Zeno paradox \cite{Misra}. In more general frameworks, we think that there exist notably suggested reasons to claim that the standard condition is satisfied. Let the open quantum system formalism be an example \cite{Dav}. There the system evolution is given by an operator semigroup which satisfies, among other properties, standarization.

\bigskip

The implications of the foregone theorem should be stated clearly. \textbf{Assumed} the description of a quantum system by a homogeneous QSES, \textbf{if} the system exhibits quantum jumps, \textbf{then} the QSES cannot be Markovian. Different attitudes can be adopted. Firstly, the possibility of mathematically representing a quantum jump through an H-valued stochastic process may be neglected, a solution we believe to be too restrictive. Secondly, it can be claimed that the quantum jump does not take place and the evolution of a quantum system, though stochastic, is continuous. This hypothesis is adopted, e.g. in the CSL theory \cite{GPR}. Nonetheless, it is also possible that the Markov condition not be satisfied even maintaining continuity, as in, e.g. \cite{Dio}. Finally, the option is left of admitting every hypothesis in the theorem with the subsequent consequences. This alternative has not been studied profoundly yet.

\end{document}